\documentclass{article}

\PassOptionsToPackage{sort, numbers}{natbib}
    
\usepackage[final]{neurips_2021}

\usepackage[utf8]{inputenc} 
\usepackage[T1]{fontenc}    
\usepackage{hyperref}       
\usepackage{url}            
\usepackage{booktabs}       
\usepackage{amsfonts}       
\usepackage{nicefrac}       
\usepackage{microtype}      
\usepackage{xcolor}         
\usepackage{graphicx}
\usepackage{caption}
\usepackage{subcaption}

\title{Evaluating saliency methods on artificial data \\with different background types}

\author{%
  Céline Budding\thanks{Work conducted when affiliated with the Technical University of Berlin and Bernstein Center for Computational Neuroscience Berlin as a MSc student.} \\
  Department of Industrial Engineering \& Innovation Sciences.\\
  Eindhoven University of Technology, Eindhoven, The Netherlands \\
  \texttt{c.e.budding@tue.nl} \\
   \And
   Fabian Eitel \\
  Charit\'e -- Universit\"atsmedizin Berlin, Berlin, Germany;\\
Bernstein Center for Computational Neuroscience Berlin, Berlin, Germany\\
   \texttt{fabian.eitel@charite.de} \\
   \AND
   Kerstin Ritter \\
    Charit\'e -- Universit\"atsmedizin Berlin, Berlin, Germany;\\
Bernstein Center for Computational Neuroscience Berlin, Berlin, Germany\\

   \texttt{kerstin.ritter@charite.de} \\
   \And
   Stefan Haufe \\
   Technische Universit\"at Berlin, Berlin, Germany;\\
   Physikalisch-Technische Bundesanstalt Braunschweig und Berlin, Berlin, Germany;\\
   Charit\'e -- Universit\"atsmedizin Berlin, Berlin, Germany;\\
   Bernstein Center for Computational Neuroscience Berlin, Berlin, Germany\\
   \texttt{haufe@tu-berlin.de} \\
}

\begin{document}

\maketitle

\begin{abstract}
    Over the last years, many `explainable artificial intelligence' (xAI) approaches have been developed, but these have not always been objectively evaluated. To evaluate the quality of heatmaps generated by various saliency methods, we developed a framework to generate artificial data with synthetic lesions and a known ground truth map. Using this framework, we evaluated two data sets with different backgrounds, Perlin noise and 2D brain MRI slices, and found that the heatmaps vary strongly between saliency methods and backgrounds. We strongly encourage further evaluation of saliency maps and xAI methods using this framework before applying these in clinical or other safety-critical settings.
\end{abstract}

\section{Introduction}
\label{introduction}
The recent increasing popularity and performance of deep neural networks (DNNs) in various fields \citep{lecun2015deep, mnih2015human, senior2020improved} has gone hand in hand with a call for artificial intelligence (AI) to be `explainable' \citep{arrieta2020explainable}. DNNs have been described as `black boxes': even though their weights and parameters are known, these are so high-dimensional that they cannot easily be understood or interpreted from a human perspective \citep{castelvecchi2016can}. Contributions in the field of explainable AI (xAI) have either called for `inherently interpretable' models \citep{rudin2019stop} or methods to provide `post-hoc interpretation' for trained DNNs \citep{doshi2017towards}. In this paper, the focus is on the latter, specifically so-called saliency methods for convolutional neural networks (CNNs) applied to image data \citep{montavon2017explaining}. Saliency methods are examples of \textit{local} methods that generate an `explanation' for each input sample. For image data, this leads to a heatmap quantifying the relevance of each pixel towards the predicted output according to some internal metric \citep{kindermans2017learning}. The saliency methods we included here are Gradient analysis \citep{baehrens2010explain, simonyan2013deep}, Layerwise Relevance Propagation (LRP, $z-$ and $\alpha\beta$-rules) \citep{bach2015pixel}, deep Taylor decomposition (DTD) \citep{montavon2017explaining}, Guided Backpropagation \citep{springenberg2014striving}, DeConvNet \citep{zeiler2014visualizing}, PatternNet \citep{kindermans2017learning} and PatternAttribution \citep{kindermans2017learning}. 

Interpretability is particularly important in safety-critical situations such as the medical field \citep[e.g.,][]{haufe2014interpretation}. Saliency methods have been applied to explain CNN decisions in multiple sclerosis \citep[MS,][]{eitel2019uncovering} and Alzheimer's disease \citep{bohle2019layer} based on magnetic resonance imaging (MRI) data. Nevertheless, these methods suffer from various problems, such as a lack of robustness \citep{eitel2019testing} and  sensitivity to low-level features \citep{sixt2020explanations}. Empirical validation schemes have been proposed \citep{samek2016evaluating, hooker2018benchmark, yang2019benchmarking}, but these only use indirect measures to determine the quality of the heatmaps. Here, we generate artificial data with different backgrounds and a known ground truth based on artificial lesions. This approach enables us to  objectively evaluate heatmaps using  quantitative metrics of \emph{explanation performance}. 

\section{Data \& Methods}
\label{data_methods}

\paragraph*{Data}
We generated artificial data consisting of a noise background multiplied with a lesion map and created two classes: Class 1, without any lesions, and class 2, with two small circular lesions with 30\% intensity and a 10 pixel diameter. Lesions were generated with a Hamming window function and constrained to the top half of the image, within the brain mask. In line with suggestions in \citep{tjoa2020quantifying}, we tested two different backgrounds: 1) Perlin noise, a type of gradient noise, with a resolution of (5, 8) pixels and 2) structural MRI data obtained from the Human Connectome Project data set\footnote{Anonymized, consent for sharing for research purposes.} \citep{van2013wu}. The 3D MRI data were preprocessed using the FMRIB Software library \citep{smith2004advances}, Then, random axial slices were selected, while discarding outer slices without brain tissue. For both background conditions, the image size was 140 x 192 and the background was normalized to $(0, 1)$ before multiplication with the lesion maps. Brain masks extracted from the MRI data set were applied to the Perlin backgrounds and lesion masks were identical across both background conditions in order to minimize variation between the sets. The samples were divided into a training (42,000 instances), a validation (6,000 instances), and a holdout set (12,000 instances). All code for this work is available here\footnote{\url{https://github.com/cebudding/MedNeurips_2021}}.

\paragraph*{Networks \& training}
\label{networks_training}
We used a VGG-inspired network containing four convolutional blocks with two convolutional layers and ReLU nonlinearity, followed by a Max Pooling layer. The number of filters (size (3, 3)) increases with each convolutional block (32, 64, 128, 256), respectively. The final two layers of the network are fully-connected layers consisting of 128 nodes with ReLU activation and two nodes with softmax activation, respectively. All nodes were initialized with a He initialization \citep{he2015delving}. The network was implemented in Keras version 2.2.4 \citep{chollet2015keras}, running on a Tensorflow version 1.12.0 backend \citep[both Apache License 2.0]{tensorflow2015-whitepaper}. Training was performed using stochastic gradient descent and categorical cross-entropy loss. All experiments were carried out on a single internal 1080TI GPU. For both datasets, the network was trained three times and the model performing best on the holdout set was used for further analysis. No extensive hyperparameter search was performed as the only aim was to obtain a network performing at around $\geq 85\%$ accuracy or better. Training continued for 125 epochs or until a loss below 0.05 was reached with a learning rate $\eta = 5e-5$ (momentum 0.9), early stopping (patience = 10 epochs, min $\delta = 5-4$), and batch size $b = 128$ with shuffling. During validation, the batch size was $b = 32$ without shuffling.

\paragraph*{Saliency methods \& evaluation}
All saliency methods were implemented through the \textit{innvestigate} package \citep[BSD License, ][]{alber2019innvestigate}. The aim was to evaluate the general usability of these methods, so they were applied using the default parameters,except for LRP-$\alpha\beta$, for which we set $\alpha = 2$ and $\beta = 1$ \citep{samek2016evaluating}. Saliency maps were computed for 200 samples from the holdout set. To evaluate the quality of the heatmaps, metrics of \emph{explanation performance} were assessed by comparing the continuous heatmaps with the binarized ground truth, i.e., inverted lesion map. ROC-AUC was included as a general performance metric, although this is biased on imbalanced datasets \citep{saito2015precision}. To reduce such bias, the mean average precision (mAP), which only focuses on the pixels in the positive, minority, class, and the precision at 99\% specificity (PREC99), which sets a high threshold such that most negative pixels are not assigned relevance, were also included. In addition to the comparison of explanation performance across methods and datasets, individual heatmaps of saliency methods were also inspected qualitatively. 

\section{Results}
\label{results}
\begin{figure}[h!]
  \centering
  \includegraphics[width=0.9\textwidth]{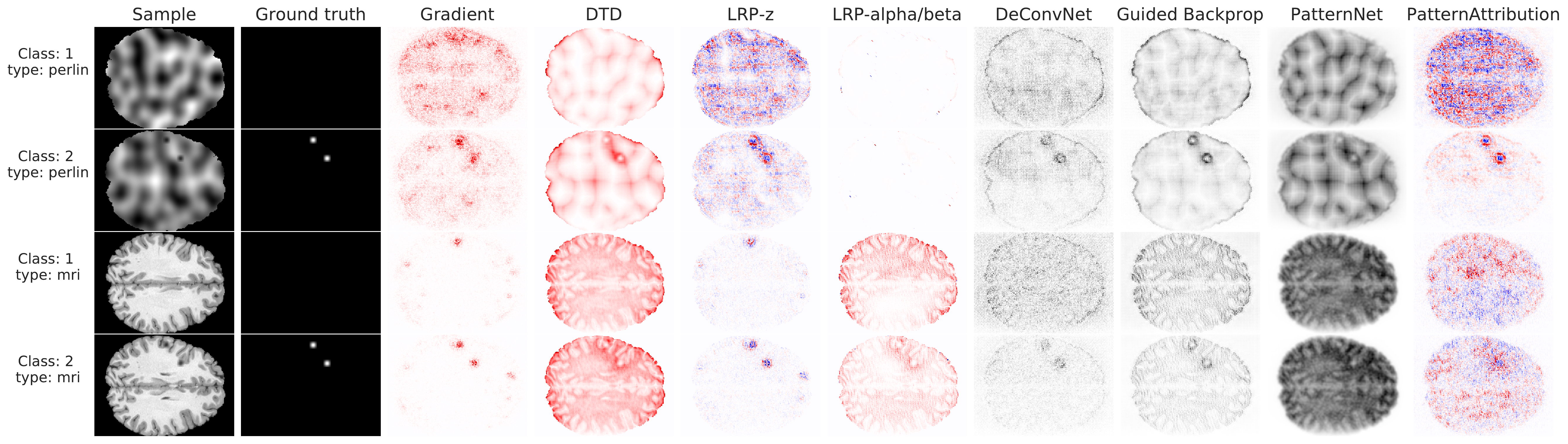}
  \caption{Example data instances and saliency maps. The first two rows show data with a Perlin background for class 1 and 2 respectively, the second two rows for data with a structural MRI background, class 1 and 2 respectively. The lesions were identical across both background conditions.}
  \label{fig:comparison_backgrounds}
\end{figure}

The average holdout set classification accuracy was $98.27 \pm 0.63\%$ for the Perlin noise data set and $94.21 \pm 3.54\%$ for the MRI data set, with a slightly higher standard deviation across three trials for the MRI data. Figure \ref{fig:comparison_backgrounds} shows  representative data instances for both background conditions, two samples from class 1 and two from class 2. The second column shows the ground truth and the rest of the columns show heatmaps from the respective saliency methods. Although only a limited number of samples is shown, some observations can be made. 
Firstly, the quality of the heatmaps seems to strongly vary across the saliency methods: whereas Gradient and LRP-$z$ seem to generate mostly relatively sharp heatmaps, methods such as DTD, Guided Backprop and PatternNet seem strongly influenced by the background.  Interestingly, Gradient, LRP-$z$, and PatternAttribution indicate spurious, non-existing lesions for class 1, mostly for the MRI background. Secondly, differences are observed between background conditions: for Gradient and LRP-$z$, heatmaps are clearer and sparser for the MRI background condition, whereas Guided Backprop and PatternAttribution seem to provide clearer heatmaps for the Perlin noise condition. As all other parameters were kept identical across classes, it is apparent that the background has an effect on the quality of the heatmaps. 

\begin{figure}
    \begin{subfigure}[b]{\textwidth}
        \centering
        \includegraphics[width=0.85\textwidth]{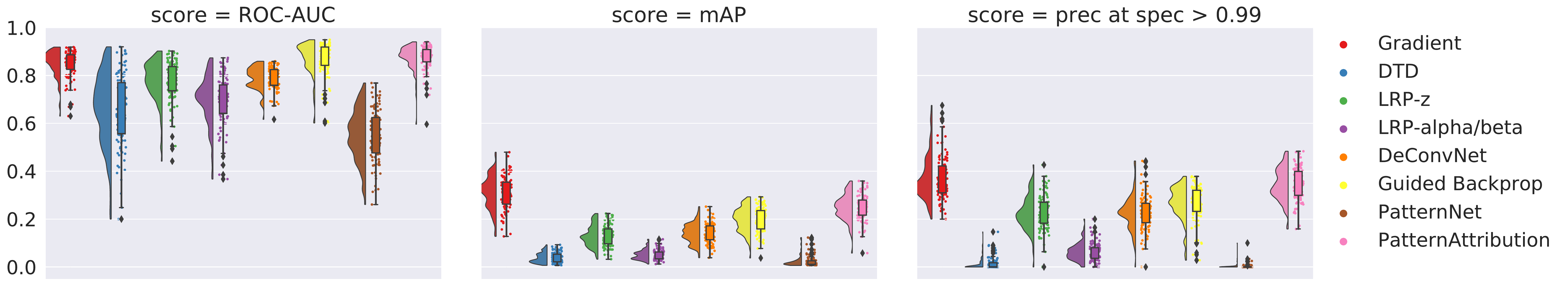}
        \caption{Performance scores (ROC-AUC, mAP, and PREC99) for a perlin noise background}
    \end{subfigure}
    \hfill
    \begin{subfigure}[b]{\textwidth}
        \centering
        \includegraphics[width=0.85\textwidth]{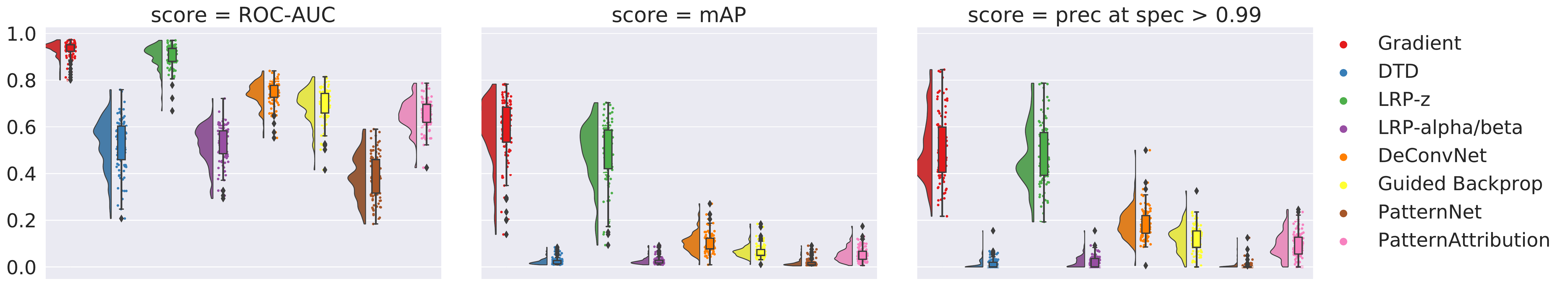}
        \caption{Performance scores (ROC-AUC, mAP, and PREC99) for an MRI background.}
    \end{subfigure}
    \caption{Explanation performance for both data sets, computed over 200 samples from class 2.}
    \label{fig:performance scores}
\end{figure}

In addition to the qualitative analysis, explanation performance metrics were calculated, see Figure \ref{fig:performance scores}. These metrics were only computed over 200 instances from the positive class (class 2) as they require binary ground truth maps in which \emph{both} types of signals (that is, lesion and background pixels) are represented. Figure \ref{fig:performance scores} suggests that ROC-AUC is biased: although the overall pattern is similar to the other metrics, almost all methods are assigned relatively high scores, which does not seem to correspond to the qualitative analysis. Therefore, mAP and PREC99 might be more reliable metrics: both show relatively high performance for Gradient in both background conditions, which corresponds to the qualitative analysis. Furthermore, LRP-$z$ seems to perform well for the MRI background, but less for the Perlin noise background.  In contrast, in line with the qualitative analysis, PatternNet performs better for the Perlin noise than for the MRI background condition. 

\section{Discussion \& Conclusion}
In this work, we present a new way to objectively evaluate heatmaps from saliency methods using artificial data with known lesions and available ground truth maps. We use this framework to generate two very similar data sets with different background conditions, Perlin noise and MR images. We show here that the background has a strong effect on the quality of the heatmaps, which is in line with \citep{tjoa2020quantifying} and might be explained by a high sensitivity to low-level features \citep{sixt2020explanations, nie2018theoretical}: for some saliency methods such as Guided Backprop, DeConvNet, and PatternNet, the background is mostly reconstructed in the saliency maps, whereas these would be expected to only indicate the lesions. Furthermore, for Gradient and LRP-$z$, heatmaps are much clearer and sparser for the MRI condition, whereas this is reversed for PatternAttribution. Based on these results and other preliminary (unpublished) results in our lab, we call for further experiments using this framework, for example investigating different noise backgrounds, such as parametric pink noise, and classification problems, for example with lesions in both classes. This might help further evaluating saliency methods and their applicability to xAI in general, and for classification of medical images in particular.

\section{Potential negative societal impact}
Saliency methods are commonly advocated to make image classification more transparent and have been applied in natural image classification \citep{lapuschkin2019unmasking} as well as for medical imaging \citep{bohle2019layer, eitel2019uncovering}. Nevertheless, the correctness and validity of the obtained heatmaps has not been sufficiently evaluated using ground truth data. In this work, we provide a framework to aid in more objective evaluation of these methods. Our work can contribute to a safer use of AI by pointing out unexpected and potentially incorrect behavior of saliency methods. As such, we expect our work to have a rather positive societal impact, and we are not aware of any potential negative impact.

\begin{ack}
This result is part of a project that has received funding from the European Research Council (ERC) under the European Union’s Horizon 2020 research and innovation programme (Grant agreement No. 758985). In addition, we acknowledge support from the German Research Foundation (DFG, 389563835; 402170461-TRR 265; 414984028-CRC 1404) and the Brain \& Behavior Research Foundation (NARSAD grant). 
The authors declare no competing interests.
\end{ack}

\bibliographystyle{apalike}
\bibliography{bibliography}

\end{document}